\documentclass[aps,preprint,showpacs,showkeys,nofootinbib,floatfix]{revtex4}
\usepackage{epsfig}
\usepackage{amssymb}
\usepackage{graphicx}
\begin{document}
\title{Hadronic resonances enhanced by thresholds}
\author{T.~F.~Caram\'es}
\affiliation{Departamento de F{\'\i}sica Fundamental and IUFFyM,
Universidad de Salamanca, 37008 Salamanca, Spain}
\author{A.~Valcarce}
\affiliation{Departamento de F{\'\i}sica Fundamental and IUFFyM,
Universidad de Salamanca, 37008 Salamanca, Spain}
\date{\emph{Version of }\today}

\begin{abstract}
We present a neat example of a meson--baryon system where the vicinity
of two different thresholds enhances the binding of a hadronic resonance,
a pentaquark. As a consequence the pattern of states may change when moving among different 
flavor sectors, what poses a warning on naive extrapolations
to heavy flavor sectors based on systematic expansions. For this purpose we simultaneously analyze 
the $N\bar D$ and $NB$ two-hadron systems looking for possible bound 
states or resonances. When a resonance is controlled by a 
coupled-channel effect, going to a different flavor sector may 
enhance or diminish the binding. This effect may, for example, generate
significant differences between the charmonium and bottomonium spectra
above open-flavor thresholds or pentaquark states in the open-charm and
open-bottom sectors.
\end{abstract}

\pacs{14.40.Lb,12.39.Pn,12.40.-y}
\maketitle

The interpretation of the recently discovered baryonic states at the LHCb, $P_c(4380)^+$ and $P_c(4450)^+$~\cite{Aai15},
as well as some of the exotic mesonic states discovered in the hidden-charm or
hidden-beauty sector is still puzzling~\cite{Bur15,Bra13}. A common feature of
all these states seems to be the proximity of their masses to two-hadron thresholds.
Their naive description as simple baryon-meson or meson-meson resonances gave
rise to predictions of bound states in heavier flavor sectors by different
spectroscopic models, like those based on the traditional meson theory of 
the nuclear forces or resorting to heavy quark symmetry arguments~\cite{Tor92,Sun11}.

In a recent paper~\cite{Vij15} a mechanism to explain the stability
or metastability of the exotic mesonic states discovered in the hidden-charm or
hidden-beauty sector was proposed. It was pointed out how two effects have to come together to allow for 
the formation of a bound state above open-flavor thresholds: the presence of two nearby thresholds and a strong 
coupling between them, in spite of the fact that the diagonal interactions 
contributing to this state are not strong. 
Thus, such mechanism, as long as it is possible avoids the risk of proliferation
of states appearing in some quark-model calculations. 

For the $X(3872)$ this is well plausible~\cite{Car09}. It was pointed out that two of the possible 
dissociation thresholds are almost exactly degenerate, the one corresponding to 
spin-singlet charmed meson plus a spin-triplet anti-charmed meson (or conjugate), 
and the one made of a light vector-meson and a charmonium vector-meson. 
For instance, in the flavor-SU(3) limit, the $H$ dibaryon 
benefits of the degeneracy of the $\Lambda\Lambda$ and $N\Xi$ thresholds, and is 
found stable in some model calculations, whereas for broken SU(3) the degeneracy 
is lost and the $H$ dibaryon becomes unstable in the same models~\cite{Ino12}. 

This idea had been already anticipated in a qualitative analysis of the possible dissociation 
thresholds of four-quark systems with a $Q\bar Q n\bar n$ structure (in the following $n$ 
stands for a light quark and $Q$ for a heavy $c$ or $b$ quark), making stringent
predictions as the non-existence of a bottom partner for the $X(3872)$ or the existence
of exotic doubly heavy mesons~\cite{Car12}. While the first prediction seems to survive experiment in
contrast to those of other theoretical models~\cite{Tor92,Sun11}, the second is still 
awaiting for an experimental effort~\cite{Hyo13}.
We wonder if there could be a neat example where one could think of some 
degeneracy of two baryon-meson thresholds leading to exotic or crypto-exotic
baryons, as it may be the case for some of the exotic meson states~\cite{Car12}.
The existence of an exotic state in a given flavor sector
can not be naively generalized to other flavor sectors in case of loss 
of the vicinity of the thresholds. In a similar manner, its non-existence
in a particular flavor sector does not exclude its presence
in different flavor sectors.

In this letter we discuss a relevant example of a five-quark state in the $NB$
two-hadron system that clearly exemplifies the importance of the mechanism we 
have previously presented. It should be considered 
in the phenomenological analysis of the recently reported pentaquark states
and may serve as a guideline for the study of the pattern of exotic states
in the baryon and meson sectors~\cite{Bur15}. Our findings come up
in the shadow of a previous study of a different two-hadron system,
$N \bar D$, whose generalization to the bottom case gave rise to an a priori
unexpected result that remarks the effect of the almost degeneracy of two 
different baryon-meson thresholds.

In Ref.~\cite{Caa12} we studied the $N \bar D$ system by means of a 
chiral constituent quark model. Our main motivation at that time was 
the study of the interaction of $D$ mesons with nucleons which is a goal of the
$\bar {\rm P}$ANDA Collaboration at the European facility FAIR~\cite{Pan09}. Thus, our
theoretical study was a challenge to be tested at the future experiments.
In this letter we perform a parallel study of the $NB$ system (with a similar 
quark structure, $nnnn\bar Q$) looking for similarities and differences with 
respect to the $N\bar D$ system. Our objective is to highlight
a particular case where the vicinity of thresholds will enhance the binding of
the baryon-meson system disrupting the number and the ordering of 
states obtained in the charm sector. Although the conclusions of this study aim to be independent 
of the particular details of the interacting model used, we for instance made use
of the chiral constituent quark model (CCQM) of Ref.~\cite{Vij05}.
It was proposed in the early 90's in an attempt to obtain a simultaneous description of the 
nucleon-nucleon interaction and the baryon spectra~\cite{Val05}. It was later on generalized to all 
flavor sectors giving a reasonable description of the meson and baryon spectra. 
The model is based on the assumption that the light--quark constituent mass appears because of the 
spontaneous breaking of the original $SU(3)_{L}\otimes SU(3)_{R}$ 
chiral symmetry at some momentum scale. In this domain of momenta, quarks interact through Goldstone 
boson exchange potentials. QCD perturbative effects are taken into account through the 
one-gluon-exchange potential. Finally, it incorporates confinement as dictated by unquenched 
lattice QCD calculations. A detailed discussion of the model can be found in Refs.~\cite{Vij05,Val05}.

The systems under study, $N \bar D$ and $NB$, do not present quark-antiquark annihilation
complications that may obscure the predictions of a particular model under some non-considered
dynamical effects. They contain a heavy antiquark, what makes the interaction 
rather simple. The quark-model used provides parameter-free predictions for 
the interaction in a baryon-meson system with charm $-1$ or bottom $+1$.
Besides, the existence of identical light quarks in the two hadrons generates 
quark-Pauli effects in some particular channels, what gives rise to an important
short-range repulsion due to lacking degrees of freedom to accommodate the 
light quarks~\cite{Val05}.

To study the possible existence of exotic states made of a light baryon, $N$ and $\Delta$,
and a charmed meson, $\bar D$ and $\bar D^*$, or a bottom meson, $B$ and $B^*$, we solve 
the Lippmann-Schwinger equation for negative energies looking at the Fredholm 
determinant $D_F(E)$ at zero energy~\cite{Gar87}. If there are no interactions then $D_F(0)=1$, 
if the system is attractive then $D_F(0)<1$, and if a
bound state exists then $D_F(0)<0$. This method permitted 
us to obtain robust predictions even for zero-energy bound states, and gave
information about attractive channels that may lodge a resonance~\cite{Car09}.
We consider a baryon-meson system $Q_i R_j$ ($Q_i=N$ or $\Delta$ and $R_j=\bar D$ 
or $\bar D^*$ for charm $-1$ and $R_j=B$ or $B^*$ for bottom $+1$) in a relative $S$ state 
interacting through a potential $V$ that contains a
tensor force. Then, in general, there is a coupling to the 
$Q_i R_j$ $D$ wave. Moreover, the baryon-meson system can couple to other 
baryon-meson states, $Q_k R_m$. We show in Table~\ref{tab1} the coupled channels in 
the isospin-spin $(T,J)$ basis for the $NB$ system (for the $N\bar D$ system one would replace
$B$ by $\bar D$ and $B^*$ by $\bar D^*$).
\begin{table}[b]
\caption{Interacting baryon-meson channels in the isospin-spin $(T,J$) basis.}
\label{tab1}
\begin{tabular}{cccc}
\hline \hline
& $T=0$  &  $T=1$ & $T=2$ \\
\hline
$J=1/2$ & $N B  - N B^* $         & $N B - N B^* - \Delta B^* $         & $\Delta B^*$ \\
$J=3/2$ & $N B^* $                    & $N B^* - \Delta B - \Delta B^* $  & $\Delta B  - \Delta B^*$ \\
$J=5/2$ &      $-$          & $\Delta B^* $                               & $\Delta B^* $ \\ \hline\hline
\end{tabular}
\end{table}
Let us briefly sketch the method to look for bound state solutions using the
Fredholm determinant. If we denote the different baryon-meson systems as channel $Q_i R_j\equiv A_n$,
the Lippmann-Schwinger equation for the baryon-meson scattering becomes
\begin{eqnarray}
t_{\alpha\beta;TJ}^{\ell_\alpha s_\alpha, \ell_\beta s_\beta}(p_\alpha,p_\beta;E)& = & 
V_{\alpha\beta;TJ}^{\ell_\alpha s_\alpha, \ell_\beta s_\beta}(p_\alpha,p_\beta)+
\sum_{\gamma=A_1,A_2,\cdots}\sum_{\ell_\gamma=0,2} 
\int_0^\infty p_\gamma^2 dp_\gamma V_{\alpha\gamma;TJ}^{\ell_\alpha s_\alpha, \ell_\gamma s_\gamma}
(p_\alpha,p_\gamma) \nonumber \\
& \times& \, G_\gamma(E;p_\gamma)
t_{\gamma\beta;TJ}^{\ell_\gamma s_\gamma, \ell_\beta s_\beta}
(p_\gamma,p_\beta;E) \,\,\,\, , \, \alpha,\beta=A_1,A_2,\cdots \,\, ,
\label{eq0}
\end{eqnarray}
where $t$ is the two-body scattering amplitude, $T$, $J$, and $E$ are the
isospin, total angular momentum and energy of the system,
$\ell_{\alpha} s_{\alpha}$, $\ell_{\gamma} s_{\gamma}$, and
$\ell_{\beta} s_{\beta }$
are the initial, intermediate, and final orbital angular momentum and spin, respectively,
 and $p_\gamma$ is the relative momentum of the
two-body system $\gamma$. The propagators $G_\gamma(E;p_\gamma)$ are given by
\begin{equation}
G_\gamma(E;p_\gamma)=\frac{2 \mu_\gamma}{k^2_\gamma-p^2_\gamma + i \epsilon} \, ,
\end{equation}
with
\begin{equation}
E=\frac{k^2_\gamma}{2 \mu_\gamma} \, ,
\end{equation}
where $\mu_\gamma$ is the reduced mass of the two-body system $\gamma$.
For bound-state problems $E < 0$ so that the singularity of the propagator
is never touched and we can forget the $i\epsilon$ in the denominator.
If we make the change of variables
\begin{equation}
p_\gamma = d\,{1+x_\gamma \over 1-x_\gamma},
\label{eq2}
\end{equation}
where $d$ is a scale parameter, and the same for $p_\alpha$ and $p_\beta$, we can
write Eq.~(\ref{eq0}) as
\begin{eqnarray}
t_{\alpha\beta;TJ}^{\ell_\alpha s_\alpha, \ell_\beta s_\beta}(x_\alpha,x_\beta;E)& = & 
V_{\alpha\beta;TJ}^{\ell_\alpha s_\alpha, \ell_\beta s_\beta}(x_\alpha,x_\beta)+
\sum_{\gamma=A_1,A_2,\cdots}\sum_{\ell_\gamma=0,2} 
\int_{-1}^1 d^2\left(1+x_\gamma \over 1-x_\gamma \right)^2 \,\, {2d \over (1-x_\gamma)^2}
dx_\gamma \nonumber \\
&\times & V_{\alpha\gamma;TJ}^{\ell_\alpha s_\alpha, \ell_\gamma s_\gamma}
(x_\alpha,x_\gamma) \, G_\gamma(E;p_\gamma) \,
t_{\gamma\beta;TJ}^{\ell_\gamma s_\gamma, \ell_\beta s_\beta}
(x_\gamma,x_\beta;E) \, .
\label{eq3}
\end{eqnarray}
We solve this equation by replacing the integral from $-1$ to $1$ by a
Gauss-Legendre quadrature which results in the set of
linear equations
\begin{equation}
\sum_{\gamma=A_1,A_2,\cdots}\sum_{\ell_\gamma=0,2}\sum_{m=1}^N
M_{\alpha\gamma;TJ}^{n \ell_\alpha s_\alpha, m \ell_\gamma s_\gamma}(E) \, 
t_{\gamma\beta;TJ}^{\ell_\gamma s_\gamma, \ell_\beta s_\beta}(x_m,x_k;E) =  
V_{\alpha\beta;TJ}^{\ell_\alpha s_\alpha, \ell_\beta s_\beta}(x_n,x_k) \, ,
\label{eq4}
\end{equation}
with
\begin{eqnarray}
M_{\alpha\gamma;TJ}^{n \ell_\alpha s_\alpha, m \ell_\gamma s_\gamma}(E)
& = & \delta_{nm}\delta_{\ell_\alpha \ell_\gamma} \delta_{s_\alpha s_\gamma}
- w_m d^2\left(1+x_m \over 1-x_m\right)^2{2d \over (1-x_m)^2} \nonumber \\
& \times & V_{\alpha\gamma;TJ}^{\ell_\alpha s_\alpha, \ell_\gamma s_\gamma}(x_n,x_m) 
\, G_\gamma(E;{p_\gamma}_m),
\label{eq5}
\end{eqnarray}
and where $w_m$ and $x_m$ are the weights and abscissas of the Gauss-Legendre
quadrature while ${p_\gamma}_m$ is obtained by putting
$x_\gamma=x_m$ in Eq.~(\ref{eq2}).
If a bound state exists at an energy $E_B$, the determinant of the matrix
$M_{\alpha\gamma;TJ}^{n \ell_\alpha s_\alpha, m \ell_\gamma s_\gamma}(E_B)$ 
vanishes, i.e., $\left|M_{\alpha\gamma;TJ}(E_B)\right|=0$.
We took the scale parameter $d$ of Eq.~(\ref{eq2}) as $d=$ 3 fm$^{-1}$
and used a Gauss-Legendre quadrature with $N=$ 20 points.

Using the method described above, we have solved the coupled-channel
problems of the $N\bar D$ and $NB$ baryon-meson systems.
The two-body interactions are obtained from the CCQM of 
Refs.~\cite{Vij05,Val05} as explained in Ref.~\cite{Caa12}.
The Pauli principle at the level
of quarks plays an important role in the dynamics of these baryon-meson systems,
because states containing Pauli blocked channels, $(T,S)=(2,5/2)$, or
Pauli suppressed channels, $(T,J)=(0,1/2)$ and $(T,J)=(1,1/2)$, are 
repulsive, see Table~\ref{tab2}. 
\begin{table}[t]
\caption{Character of the interaction in the different $N\bar D$ $(T,J)$ states.}
\label{tab2}
\begin{tabular}{cccc}
\hline
\hline
& $T=0$  &  $T=1$ & $T=2$ \\
\hline
$J=1/2$ &  Repulsive           &  Repulsive         &  Weakly repulsive      \\
$J=3/2$ &  Weakly attractive   &  Weakly repulsive  &  Attractive            \\
$J=5/2$ &  $-$                 &  Attractive        &  Strongly repulsive    \\ \hline \hline
\end{tabular}
\end{table}
\begin{figure*}[b]
\hspace{-1cm}
\resizebox{7.5cm}{7.5cm}{\includegraphics{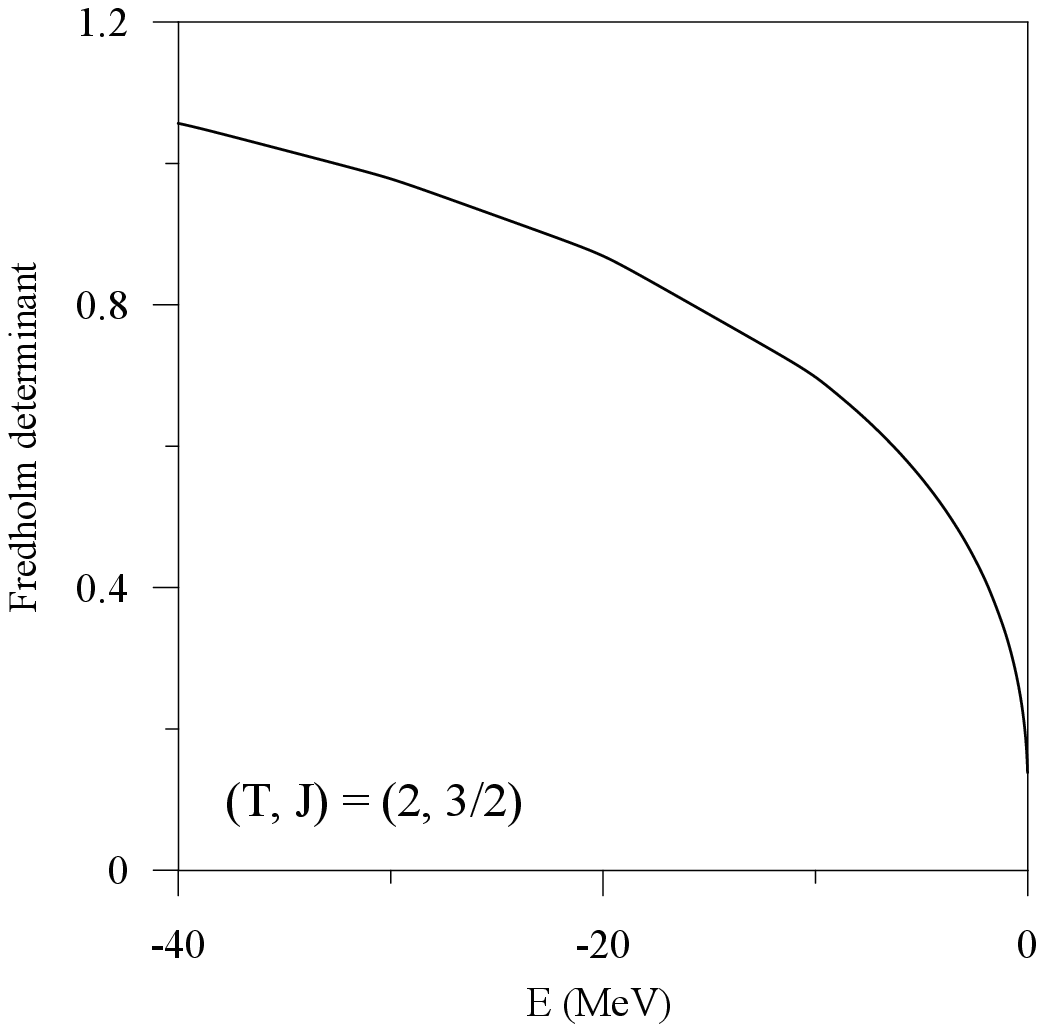}}\hspace{+1cm}
\resizebox{7.5cm}{7.5cm}{\includegraphics{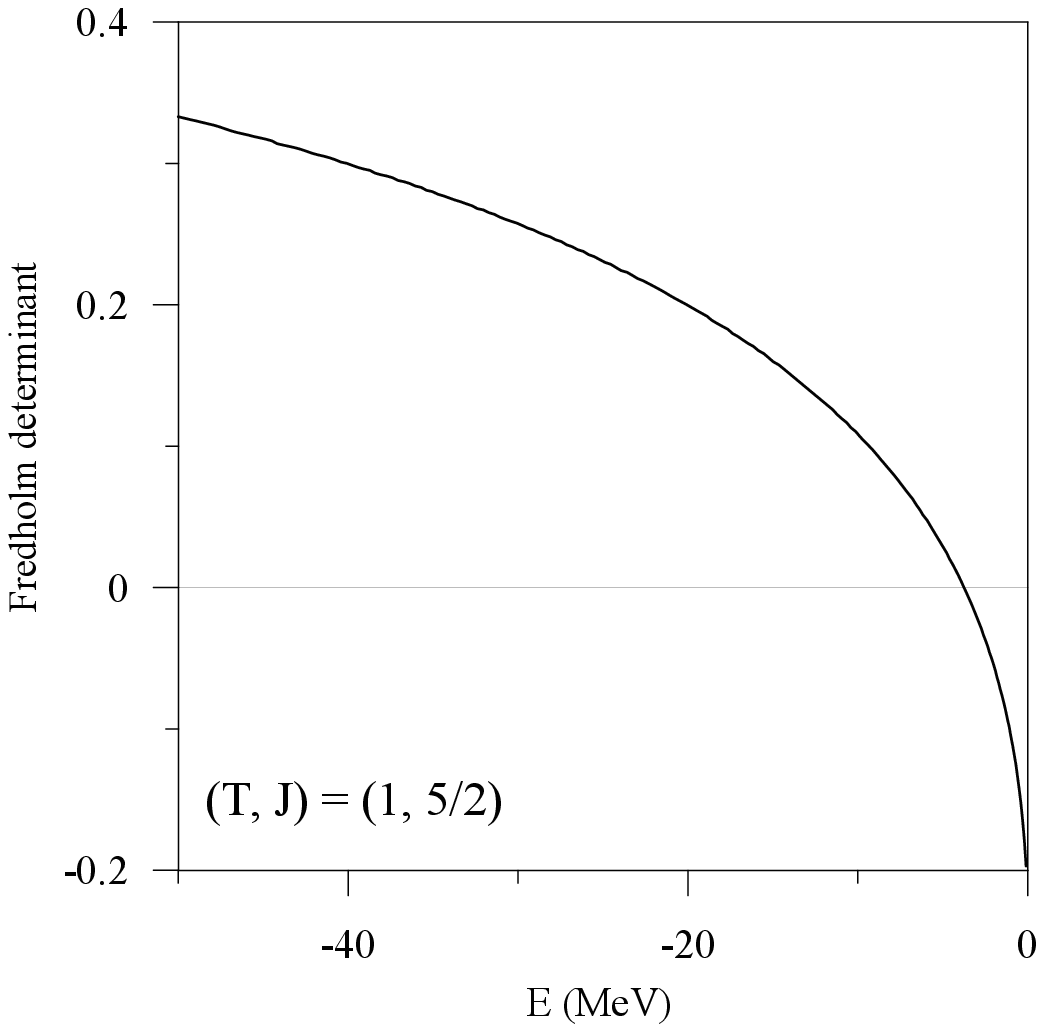}}
\caption{$(T,J)=(2,3/2)$ (left panel) and $(T,J)=(1,5/2)$ (right panel) Fredholm determinant
of the $N\bar D$ system.}
\label{fig1}
\end{figure*}

Let us first discuss the $N\bar D$ system.
In Table~\ref{tab2} we summarize the character of the interaction 
for the different $(T,J)$ states, being the most attractive ones
the $(T,J)=(2,3/2)$ and $(T,J)=(1,5/2)$. We show in Fig.~\ref{fig1} their Fredholm 
determinant. The state with quantum numbers $(T)J^P=(1)5/2^-$
shows a bound state with a binding energy of 3.9 MeV. It corresponds to a
unique physical system, $\Delta \bar D^*$, that
would appear in the scattering of $\bar D$ mesons on nucleons as a D wave resonance, what 
could in principle be measured at $\bar {\rm P}$ANDA~\cite{Pan09}. The $(T,J)=(2,3/2)$ state contains a
coupled-channel problem, $\Delta \bar D  - \Delta \bar D^*$. While the diagonal interactions
are not even attractive, the coupling between them is strong because the decay $\bar D^* \to
\bar D + \pi$ is allowed, but not enough to generate a bound state as can be seen in the
left panel of Fig.~\ref{fig1}.

\begin{figure*}[t]
\hspace{-1cm}
\resizebox{7.5cm}{7.5cm}{\includegraphics{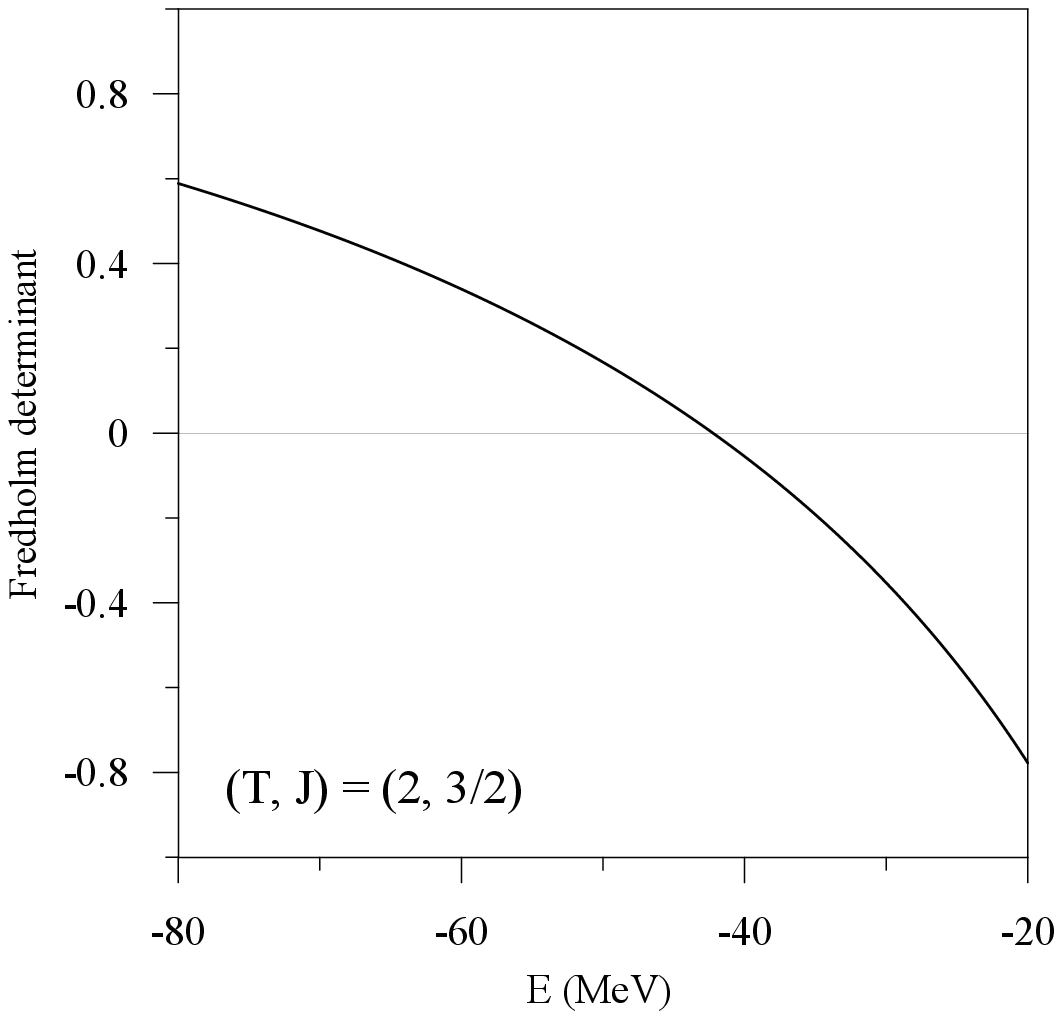}}\hspace{+1cm}
\resizebox{7.5cm}{7.5cm}{\includegraphics{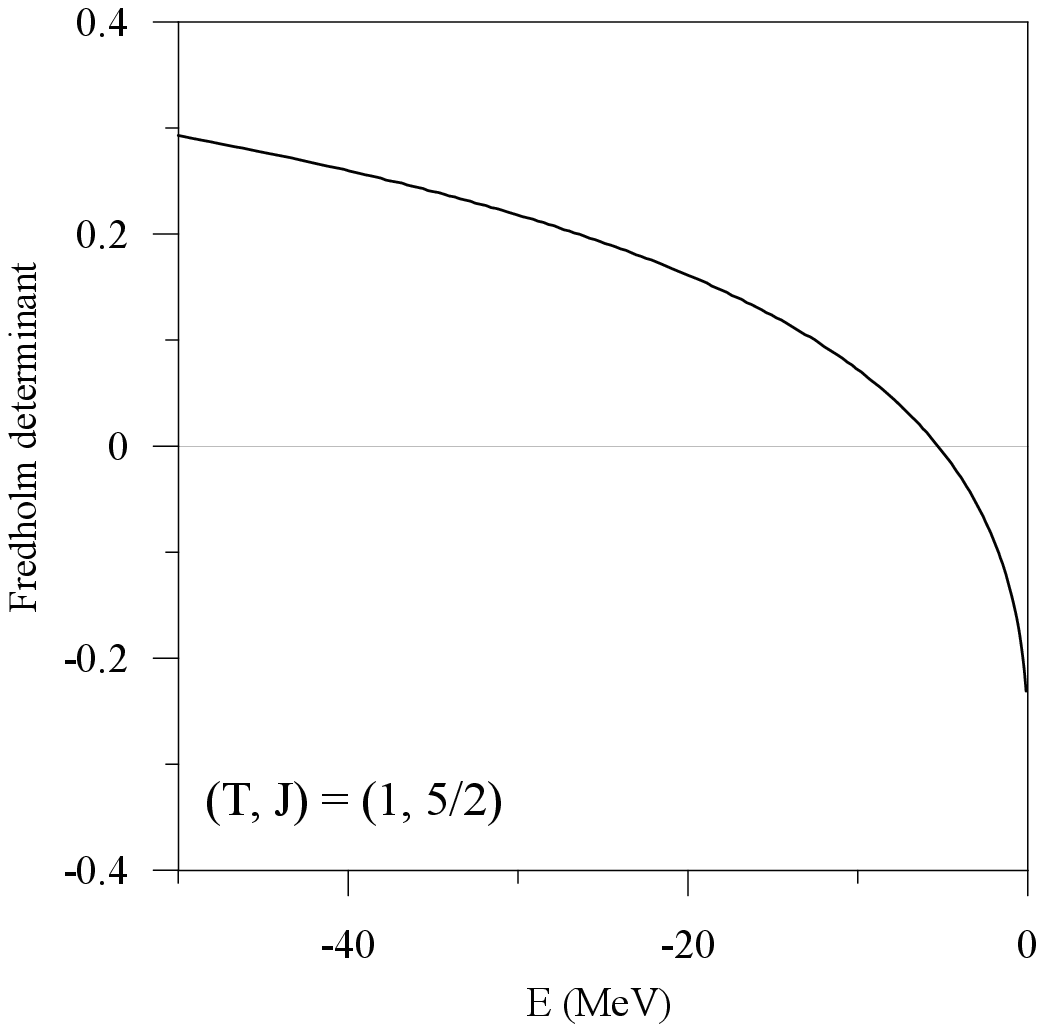}}
\caption{$(T,J)=(2,3/2)$ (left panel) and $(T,J)=(1,5/2)$ (right panel) Fredholm determinant
of the $NB$ system.}
\label{fig2}
\end{figure*}

When the mass of the heavy meson is increased the contribution of the 
kinetic energy is reduced. Thus, the binding energy of the $NB$ system 
is expected to be slightly larger than in the $N \bar D$ case, because $\bar D$ 
and $B$ mesons have similar interactions with nucleons due to having
the same quark structure.
This is why we have repeated the calculation for the $NB$ system looking
for deeper bound states in a baryon-meson system with a heavier antiquark.
We depict in Fig.~\ref{fig2} the Fredholm determinant of the two
channels that were found to be attractive in the $N \bar D$ system, see Table~\ref{tab2}.
Surprisingly, the increment in the attraction is not regularly spread over the
different $(T,J)$ channels. The $(T,J)=(2,3/2)$ is strongly affected and becomes
the lowest one with an important gain of binding, showing a bound state with 
a binding energy of 42 MeV. The ordering of 
the attractive channels is therefore reversed with respect to the $N\bar D$ system
as can be seen in Fig.~\ref{fig2}.

What is the responsible for this unexpected behavior of the binding energy as 
the mass of the heavy meson augments?
The reason lies on the internal structure of the states and the behavior of
the thresholds when increasing the heavy meson mass. As mentioned above, the $(T,J)=(1,5/2)$ state is made of
a unique physical system, $\Delta \bar D^*$ in the charm sector or $\Delta B^*$ in the bottom
sector, and thus there are no coupled-channel effects. Moving from the charm to the bottom 
sector gives rise to a small gain of binding, from 3.9 MeV in the $N\bar D$ system to
5.3 MeV in the $NB$ system, as one would naively have expected due to a
smaller kinetic energy contribution but keeping a rather similar interaction.
However, the $(T,J)=(2,3/2)$ state contains a coupled-channel 
problem, $\Delta \bar D  - \Delta \bar D^*$ in the charm sector and 
$\Delta B  - \Delta B^*$ in the bottom sector. Whereas the diagonal potentials
are not strong, as mentioned above, the coupling between the two channels 
is important because the $\bar D^* \to \bar D + \pi$ or $B^* \to B + \pi$ decays
are allowed. When moving from the charm to the bottom sector the most important effect is the
reduction of the mass difference between the two thresholds contributing to this state. 
The mass difference between vector and pseudoscalar mesons scales as predicted by
the chromomagnetic interaction~\cite{Isg99,Clo03}, $1/{m_qm_Q}$. This means a reduction around a factor
3 when going from open-charm to open-bottom mesons. In particular, while
$M(\Delta \bar D^*) - M(\Delta \bar D)=$ 141 MeV, 
$M(\Delta B^*) - M(\Delta B)=$ 45 MeV what makes the coupled-channel effect much more important
in the bottom sector and reverses the order of the two attractive channels, 
making the $(T,J)=(2,3/2)$ state the lowest one. Similarly, the $(T,J)=(0,1/2)$ channel, that is repulsive for the
$N\bar D$ system, becomes attractive in the $NB$ case due to the coupled-channel 
effect, with a resonance at threshold. Thus, when going from the
charm to the bottom sector in the baryon-meson open-flavor region, the number of states and their ordering
may be modified due to the presence of nearby thresholds. Such effect seems hardly difficult to be
predicted by any systematic expansion of the heavy quark sector.
\begin{figure}[t]
\vspace*{-1cm}
\hspace*{-1cm}\mbox{\epsfxsize=80mm\epsffile{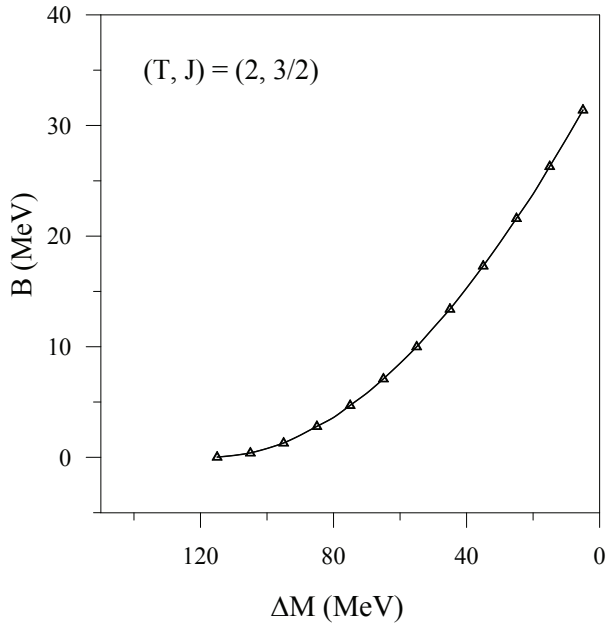}}
\caption{$(T,J)=(2,3/2)$ $N\bar D$ binding energy, B in MeV, as a function of
$\Delta M=M(\Delta \bar D^*) - M(\Delta \bar D)$ mass difference. Note that
for the experimental value, 141 MeV, the system is not bound.}
\label{fig3}
\end{figure}

Thus, as we had theoretically
predicted in our benchmark calculation of Ref.~\cite{Vij15}, one observes how two effects
come together to form a deeply bound state, the presence of two nearby thresholds and a strong
coupling between them, in spite of the fact that the diagonal interactions of the different
quantum numbers contributing to this state are not strong. 
Hence, threshold vicinity is a required but not sufficient 
condition to find a bound state. 
To illustrate our results, we have plotted in 
Fig.~\ref{fig3} the evolution of the binding energy of the $(T,J)=(2,3/2)$ $N\bar D$ state if
we artificially diminish the mass difference between the vector $\bar D^*$ and the pseudoscalar 
$\bar D$ mesons. We can see how a bound state arises when the thresholds come closer, around 120 MeV, without
modifying the interactions entering the problem. Besides, the binding energy increases when the mass difference
is reduced. This result poses an important warning when trying to extrapolate results of
binding energies of two-hadron systems to different flavor sectors. If the binding is mainly due to the
vicinity of coupled thresholds, the increase on the mass of the two-hadron system
may diminish the binding if it separates the thresholds~\cite{Car12}. Thus, if this mechanism
is working for some of the recently discovered pentaquark states at the LHCb
or the exotic states discovered in the hidden-charm or hidden-beauty meson spectra,
the pattern expected on different flavor sectors may differ significantly
as opposite to the charmonium and bottomonium spectra or the charm and bottom
baryon spectra below open-flavor thresholds. 

Let us finally note that the one-pion exchange plays a significant role in 
the $\Delta \bar D  - \Delta \bar D^*$ and $\Delta B  - \Delta B^*$ transition 
potentials, being the only contribution if quark-exchange effects were 
neglected (see Ref.~\cite{Caa12} for details). As explained above, the mixing
induced by the one-pion exchange, as well as quark-exchange contributions coming
from the other terms of the interacting potential, comes enhanced due to the 
reduction of the mass difference between the two thresholds in the bottom sector.
It has been recently argued in Ref.~\cite{Vol16} the existence of an approximate
light quark spin symmetry in $Z_b$ resonances, implying that the part of the
interaction between heavy mesons that depends on the total spin of the light
quark and antiquark is strongly suppressed. Unfortunately, this hypothesis
cannot be tested in the system under study due to the presence of a single
heavy antiquark. This symmetry will be proved in future works of 
$Z_c$~\cite{Aai14,Chi14} and $Z_b$ resonances~\cite{Bon12} as well as
the pentaquark states reported at LHCb~\cite{Aai15}, where the mechanism
proposed in this letter will be tested.

It is interesting to note that a similar argument has been drawn in Ref.~\cite{Bar15} where the {\it supercharmonium} states,
experimental resonances that appear to contain a $c\bar c$ pair, have 
been introduced but have other properties that preclude a description in terms of only $c\bar c$ (idealized charmonium) basis states. Four-quark
supercharmonium configurations occur very near, or even below, the lowest S-wave threshold
for production of meson pairs. These states evade Coleman's argument in 1979~\cite{Col79}, who used the $1/N_c$
expansion of QCD to show that four--quark color singlets tend to propagate as pairs of mesons. One of these 
supercharmonium states could be the $Z(4475)$, that may appear as a linear superposition
of $D\bar D^*$ mesons with one of them in a radially excited state $2S$. In a simple
one-pion exchange model the $D \to D + \pi$ vertex is forbidden while the $D^* \to D + \pi$
is allowed. This gives rise to weak diagonal interactions but a strong mixing between
two nearby thresholds, $D(1S)\bar D^*(2S)$ with a mass of 4482 MeV/c$^2$, and $D^*(1S) \bar D (2S)$ with a mass of 4433 MeV/c$^2$. 
The small mass difference between the
thresholds due to the reduction of the hyperfine splitting when increasing the radial
excitation leads to an amplification of the binding.

One can find also a similar situation in the heavy baryon spectra.
An important source of attraction might be the coupled-channel effect of the two 
allowed thresholds, $(nnn)(Q\bar n)$ and $(nnQ)(n\bar n)$~\cite{Lut05}.
When the $(nnn)(Q\bar n)$ and $(nnQ)(n\bar n)$ thresholds are sufficiently 
far away, the coupled-channel effect is small, and bound states are not found. 
However, when the thresholds get closer the coupled-channel strength is increased, 
and bound states may appear for a subset of quantum numbers. Under these 
conditions, there are states with high spin $J^P=5/2^-$  that may lodge a 
compact five-quark state for all isospins~\cite{Car14}. The reason stems on the reverse of the ordering of the
thresholds, being the lowest threshold $(nnn)(Q\bar n)$ the one with the more attractive
interaction. Of particular interest is the $(T)J^P=(2)5/2^-$ state, that survives
the consideration of the break apart thresholds. It may
correspond to the $\Theta_c(3250)$ pentaquark found by the QCD sum rule analysis of Ref.~\cite{Alb13}
when studying the unexplained structure with a mass of 3250 MeV/c$^2$ in the $\Sigma_c^{++} \pi^- \pi^-$ 
invariant mass reported recently by the BABAR Collaboration~\cite{Lee12}. 
Such state could therefore be a consequence of the close-to-degeneracy of the lowest thresholds with
$(T)J^P=(2)5/2^-$, $\Delta D^*$ and $\Sigma^*_c \rho$, and the attractive interaction
of the $\Delta D^*$ system.

Summarizing, we have studied the $N \bar D$ and $NB$ two-hadron systems at low energies by means of a chiral 
constituent quark model. We have found a clear example of a baryon-meson system, $NB$, where the vicinity
of two different thresholds enhances the binding of a hadronic resonance as compared to the 
$N\bar D$ system. As a consequence the pattern of states may change when moving among different 
flavor sectors, both in number and relative ordering of states, what poses a warning on 
extrapolations to heavy flavor sectors based on systematic expansions. 
When a resonance is controlled by a coupled-channel effect, going to a different flavor sector may 
enhance or diminish the binding. This effect may, for example, generate
significant differences between the charmonium and bottomonium spectra
above open flavor thresholds or pentaquark states in the open-charm and
open-bottom sectors.

\acknowledgments 
This work has been partially funded by Ministerio de Educaci\'on y Ciencia and 
EU FEDER under Contracts No. FPA2013-47443-C2-2-P and FPA2015-69714-REDT and
by FAPESP grant 2015/50326-5.

\end{document}